\documentclass[]{article}
\usepackage[margin=3cm]{geometry}
\usepackage{color}
\usepackage{url}
\usepackage{graphicx}
\usepackage{geometry}
\usepackage{lipsum}
\usepackage[export]{adjustbox}
\usepackage[font=footnotesize,labelfont=bf]{caption}
\usepackage{mathtools} 
\usepackage[T1]{fontenc}
\usepackage{gensymb}
\usepackage{amsmath}
\usepackage{amssymb}
\usepackage{fancyhdr}
\usepackage[utf8]{inputenc}
\usepackage{authblk}

\title{Stochastic modeling of auto-regulatory genetic feedback loops: a review and comparative study}
\author{J. Holehouse, Z. Cao and R. Grima}
\affil[]{School of Biological Sciences, University of Edinburgh, UK}

\begin{document}

\maketitle
\begin{abstract}
Auto-regulatory feedback loops are one of the most common network motifs. A wide variety of stochastic models have been constructed to understand how the fluctuations in protein numbers in these loops are influenced by the kinetic parameters of the main biochemical steps. These models differ according to (i) which sub-cellular processes are explicitly modelled; (ii) the modelling methodology employed (discrete, continuous or hybrid); (iii) whether they can be analytically solved for the steady-state distribution of protein numbers. We discuss the assumptions and properties of the main models in the literature, summarize our current understanding of the relationship between them and highlight some of the insights gained through modelling.  
 
\end{abstract}

\section{Introduction}
Gene regulatory networks (GRNs) provide an abstraction of the complex biochemical interactions behind transcription and translation, the central dogma of molecular biology. Feedback has been identified as an important motif in such networks, defined through the regulation of an \textit{upstream} process by one \textit{downstream} of it. Auto-regulation is the most basic kind of feedback loop -- a protein expressed from a gene activates or suppresses its own transcription. These lead to positive or negative feedback, respectively. It has been estimated that 40\% of all transcription factors in \textit{E. coli} self-regulate \cite{rosenfeld2002negative} with most of them participating in auto-repression \cite{shen2002network}. Many biological systems utilize a combination of positive and negative feedback loops, such as the circadian and segmentation clocks \cite{takahashi2017transcriptional,wiedermann2015balance}.  

Therefore, the computational and experimental study of the behaviour of auto-regulatory feedback loops is an important field of study. Measurement of the distribution of protein numbers in living cells using fluorescence microscopy \cite{elf2019single} is now a routine process. Mathematical models represent a useful tool to understand what sort of interactions in feedback loops lead to observed protein distributions, potentially leading to insight into how noise (large fluctuations in gene products with low copy numbers \cite{elowitz2002stochastic}) is managed at the subcellular level \cite{hooshangi2006effect,singh2009optimal}. These models have also been used to understand how auto-activation influences the sensitivity to input signals and the speed of induction \cite{hermsen2011speed} and to gain insight into the sources of noise in auto-regulatory networks \cite{liu2016decomposition,jia2017stochastic}. Various inference approaches have also been devised to estimate the rate constants characterizing feedback loops from population snapshot data \cite{milner2013moment,stathopoulos2013markov,cao2019accuracy,ocal2019parameter}.

The conventional stochastic description of gene regulatory networks is given by the \textit{chemical master equation} (CME), a time-evolution equation for the probability of observing a certain number of gene products at a given time \cite{schnoerr2017approximation}. This Markovian description is discrete in the sense that it takes into account that molecule numbers change by integer amounts when a reaction occurs. In Section 2 we describe the most common coarse-grained CME models for auto-regulatory feedback loops, elucidate the relationship between them and identify the regions of parameter space where their analytical predictions for the distribution of protein numbers are accurate compared to a fine-grained model. In Section 3 we compare and discuss continuous approximations of the CME using Fokker-Planck equations and partial integro-differential equations and briefly review other continuous approaches. In Section 4 we outline the main biological insights obtained using stochastic models. Finally, we conclude in Section 5 where we identify open problems.

\section{Discrete Models of auto-regulation}

From a biologist perspective, a minimal model of auto-regulation should describe the main biochemical processes describing the flow of information from gene to mRNA to protein and back to the gene. Hence the model should describe transcription and translation (the two steps at the heart of the central dogma of molecular biology), mRNA and protein degradation, and interactions of proteins with genes. For simplicity we consider the case where there is a single gene copy and all processes are modelled as first-order reactions except the protein-gene interactions which naturally follow second-order kinetics. We refer to this model as the full model since it will be our ground truth, i.e. the finest scale model that we shall consider here. The reactions describing this model are $G \xrightarrow[]{\rho_u} G + M, M \xrightarrow[]{d_m} \varnothing, M \xrightarrow[]{k} M + P, P + G \xrightarrow[]{\sigma_b} G^*, G^* \xrightarrow[]{\sigma_u} G + P, G^* \xrightarrow[]{\rho_b} G^* + M, P \xrightarrow[]{d_p} \varnothing$. While this model is intuitive, it has not been studied extensively because the mathematical description of its stochastic dynamics, as provided by the chemical master equation, is not easy to solve analytically. In fact even in the absence of the feedback loop, i.e. no protein-gene interactions, its master equation has still not been solved exactly \cite{shahrezaei2008analytical}. Hence historically, simplified versions of the full model have received much more attention in the literature. These are the models by Hornos et al. \cite{hornos2005self} in 2005, Grima et al. \cite{grima2012steady} in 2012 and Kumar et al. \cite{kumar2014exact} in 2014. Henceforth we shall refer to these as the Hornos, Grima and Kumar models. There exist other discrete models e.g. \cite{vandecan2013self} which in certain limits reduce to the aforementioned three.

\begin{figure}[h!]
\centering
\includegraphics[width=0.9\textwidth]{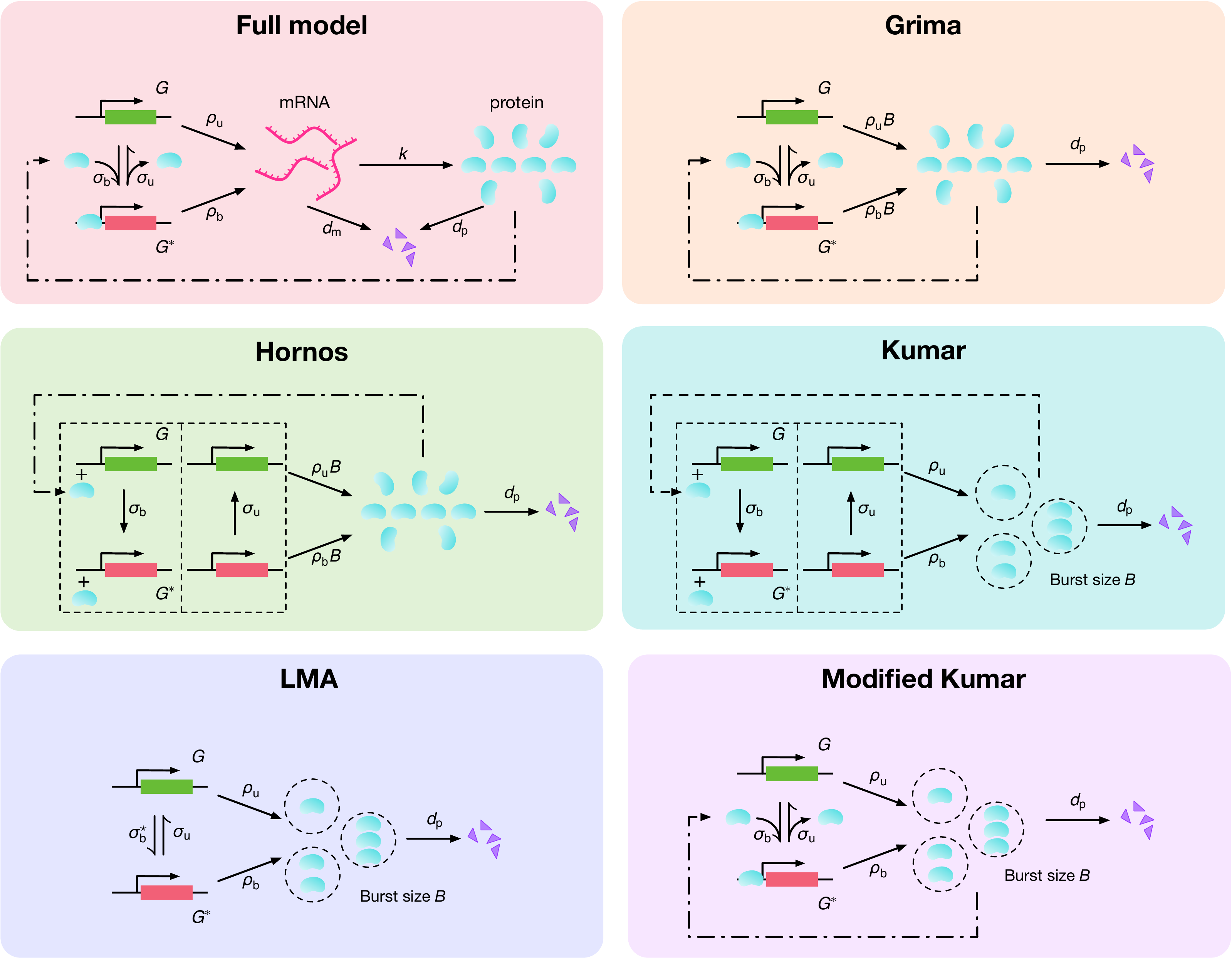}
\caption{Models of an auto-regulatory feedback loop. The full model has an explicit description of both mRNA and protein but its CME has no known exact steady-state analytical solution. The Grima, Hornos and Kumar models represent  approximations of the full model wherein only the protein is explicitly described and the CME can be solved exactly in steady-state. The Grima and Hornos models assume proteins are produced one at a time while Kumar assumes bursty production with mean burst size $B$ (denoted by dashed circles). The Hornos and Kumar models neglect protein number fluctuations due to protein-gene binding and unbinding, whereas the Grima model takes them into account. The Modified Kumar model is same as the Kumar model but takes into account fluctuations due to binding-unbinding; its CME has no known exact steady-state solution. The LMA is a discrete approximation of the full model given by the exact solution of the CME of a bursty protein production process with promoter switching and no feedback. The rate of switching to state $G^*$ is not $\sigma_b$ but $\sigma_b^*$ which is a function of all the parameters in the full model (see Supplementary Note 7 in \cite{cao2018linear}). Note that under the assumption of rapid mRNA equilibration, the mean rate of protein production is the same in all models and hence the models are indistinguishable when fluctuations are ignored.}
\label{fig1}
\end{figure}

The three models share a few common properties: (i) They only describe protein fluctuations, i.e. there is no explicit mRNA description. (ii) The models are discrete in the sense that protein numbers change by discrete integer amounts when reactions occur. (iii) The chemical master equation for each model admits an exact solution in steady-state conditions. These exact solutions have been obtained using the method of generating functions but in other studies using similar models, the solution was obtained using the Poisson representation \cite{liu2016decomposition,gardiner1985handbook,sugar2014self,iyer2014mixed}. There are however important differences between the models particularly how they describe protein production and protein-gene interactions that are not often spelled out but can be discerned from the form of the CME. The Hornos model assumes protein molecules are produced one at a time and neglects changes in the protein molecule numbers when a protein binds or unbinds from the gene. The Hornos model (excluding bound-protein degradation) is explicitly given by the set of reactions: $G \xrightarrow[]{\rho_u B} G + P, P \xrightarrow[]{d_p} \varnothing, P + G \xrightarrow[]{\sigma_b} P + G^*, G^* \xrightarrow[]{\sigma_u} G, G^* \xrightarrow[]{\rho_b B} G^* + P$. Note that the effective rate of protein production in state $G$ is $\rho_u B$ where $B = k / d_m$ (the mean number of proteins produced by an mRNA molecule during its lifetime). The reason for this is that if we define $\langle n_m \rangle$ as the mean mRNA number in the full model then the effective mean production rate is $\rho_u \langle n_m \rangle \approx \rho_u k / d_m = \rho_u B$ when mRNA equilibrates rapidly ($d_m \gg d_p$, a common assumption as we discuss later). The same analysis follows for the effective production rate in state $G^*$. The Grima model also assumes protein molecules are produced one at a time but takes into account protein fluctuations from the binding-unbinding process, i.e. when a protein binds a gene, the protein numbers are decreased by one and conversely they are increased by one when unbinding occurs. The Grima model (excluding bound-protein degradation) is explicitly given by the set of reactions: $G \xrightarrow[]{\rho_u B} G + P, P \xrightarrow[]{d_p} \varnothing, P + G \xrightarrow[]{\sigma_b} G^*, G^* \xrightarrow[]{\sigma_u} G + P, G^* \xrightarrow[]{\rho_b B} G^* + P$. The Kumar model is similar to the Hornos model except that proteins are produced in a burst (a phenomenon called translational bursting) where the burst size is a random number. Specifically, the Kumar model is given by the set of reactions: $G \xrightarrow[]{\rho_u} G + r P, P \xrightarrow[]{d_p} \varnothing, P + G \xrightarrow[]{\sigma_b} P + G^*, G^* \xrightarrow[]{\sigma_u} G, G^* \xrightarrow[]{\rho_b} G^* + r P$, where is $r$ is a positive integer drawn from the geometric distribution with mean $B$. Note that the mean rate of protein production in each of the two promoter states is the same as in the full, Grima and Hornos models. The differences between the models are illustrated in Fig. 1. 

The relationship of these models to each other and to the full model is still not completely understood. In Fig. 2 we summarize our current understanding of the regions of parameter space where the models' analytical prediction of the steady-state protein number distribution agrees with stochastic simulations of the full model (using the stochastic simulation algorithm, SSA \cite{gillespie2007stochastic}) for the case of positive feedback (panels I-IV) and negative feedback (panels V-VIII). We enforce fast promoter switching conditions by choosing the rate constants of protein-gene binding ($\sigma_b$) and unbinding ($\sigma_u$) to be large compared to the other rate constants. For the full model we also choose protein degradation rates $d_p = 1$ to be significantly smaller than mRNA degradation rates $d_m = 10$. Both of these conditions are common to many genes in both prokaryotic and eukaryotic cells \cite{shahrezaei2008analytical,sepulveda2016measurement}. For each type of feedback, there are 4 plots for combinations of small and large values of $L$ and $B$ where $L =\sigma_u / \sigma_b$ is the ratio of unbinding to binding rates (inversely proportional to the feedback strength) and $B=k/d_m$ is the mean burst size. For completeness, we also show the deterministic rate equation prediction for the protein number in the full model (vertical orange lines). 

The following considerations allow us to deduce which models are accurate in which part of parameter space. Models that assume protein fluctuations in the binding-unbinding process are negligible are incorrect when feedback is strong ($L$ is small), as recently shown in \cite{holehouse2019revisiting} -- we shall call this Property 1. Models that assume proteins are produced one molecule at a time are only correct when $B$ is small -- we shall call this Property 2. The reason for the latter property is as follows. It is well known that when mRNA decays much faster than protein then the production of proteins in the full model without feedback occurs in bursts of random size described by a geometric distribution with a mean of $B$ and the Fano factor of the protein distribution is $1 + B$ \cite{shahrezaei2008analytical}. In models that assume proteins are produced one at a time, if there was no feedback then the Fano factor of the protein distribution would be $1$ (since the distribution would be Poisson). Hence these models (with or without feedback) can only provide a good approximation to the full model when $B$ is small. Armed with Properties 1 and 2, we can now explain Fig. 2. For strong feedback (small $L$), independent of burst size, by property 1 both the Hornos and Kumar models fail to accurately match the full model. For low feedback (large $L$), the Kumar model is accurate for all burst sizes whereas the Hornos model is only accurate for small burst sizes (by Property 2). The Grima model is only accurate for low burst sizes (by Property 2) independent of feedback strength. It is noteworthy that by modifying the Kumar model so that it takes into account bursting (illustrated in Fig. 1 as Modified Kumar) then the steady-state protein distributions obtained from the SSA of this model are practically indistinguishable from the SSA of the full model. It is currently unknown if this model admits an exact analytical steady-state solution. 

\begin{figure}[h!]
\centering
\includegraphics[width=1\textwidth]{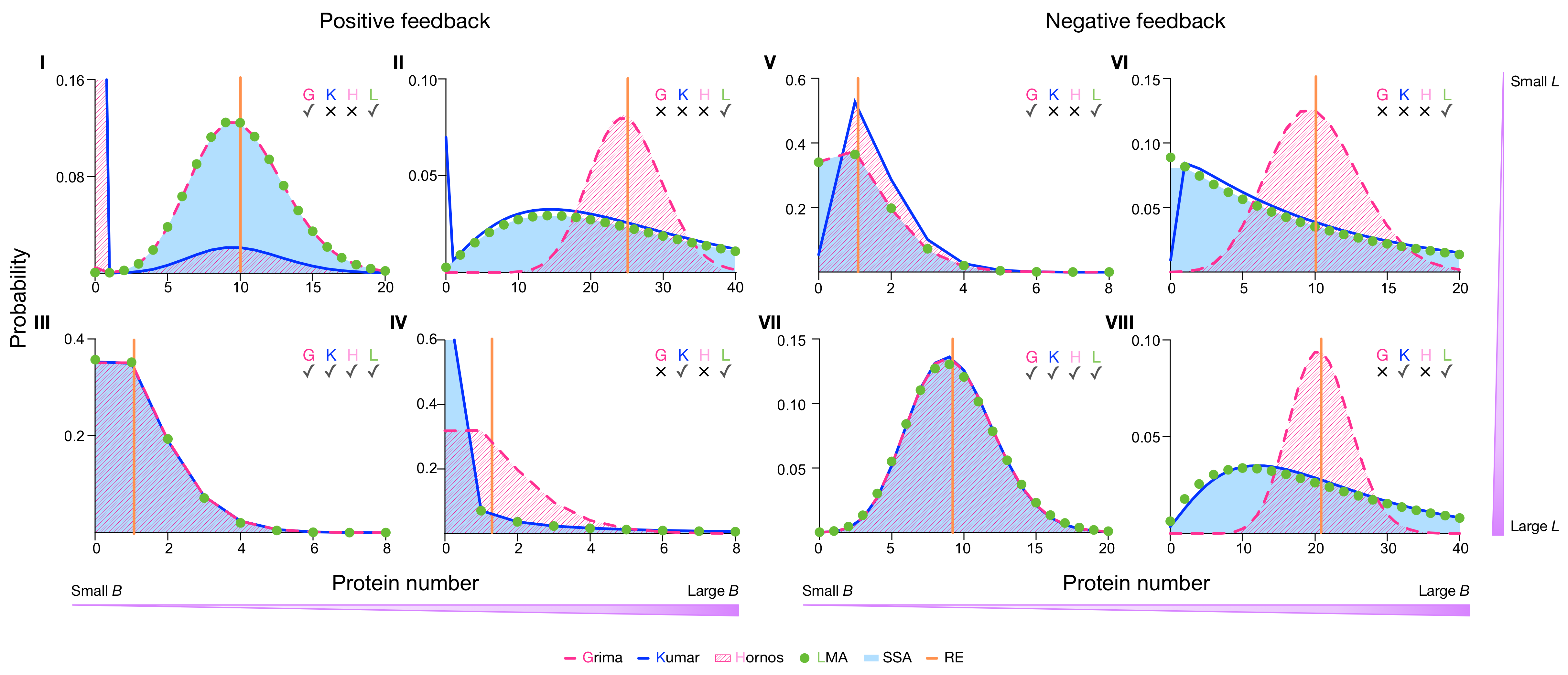}
\caption{Comparison of stochastic simulations of the full model (denoted as SSA) with the steady-state analytical distributions predicted by the reduced models of Grima (G), Hornos (H), Kumar (K) and the LMA (illustrated in Fig. 1). Panels I-IV show positive feedback ($\rho_u < \rho_b$) and Panels V-VIII show negative feedback ($\rho_u > \rho_b$). The rate equation prediction for mean protein number in the full model is denoted as RE. Note that $L=\sigma_u/\sigma_b$ is inversely proportional to the feedback strength and $B=k/d_m$ is the mean protein burst size. The LMA provides the most accurate discrete approximation of the full model (8 out of 8 regions of parameter space), followed by Grima/Kumar (4 out of 8 regions) and Hornos (2 out of 8 regions). See text for discussion. The parameters for small $L$ (I, II, V, VI) are $\sigma_u = 10^3, \sigma_b = 10^5$ and for large $L$ (III, IV, VII, VIII), they are $\sigma_u = 10^5, \sigma_b = 10^3$. The parameter for small $B$ (I, III, V, VII) is $B = 10^{-2}$ and for large $B$ (II, IV, VI, VIII), it is $B = 10$. The decay rates are fixed to $d_m = 10, d_p = 1$. The rest of the parameters are: (I) $\rho_u=10^{-2}$, $\rho_b=10^{3}$; (II) $\rho_u=10^{-1}$, $\rho_b=2.5$; (III) $\rho_u=10^{2}$, $\rho_b=10^{3}$; (IV) $\rho_u=10^{-1}$, $\rho_b=2.5$; (V) $\rho_u=10^{3}$, $\rho_b=10^{2}$; (VI) $\rho_u=10$, $\rho_b=1$; (VII) $\rho_u=10^{3}$, $\rho_b=10^{2}$; (VIII) $\rho_u=2.5$, $\rho_b=10^{-1}$.}
\label{fig1}
\end{figure} 
 
Another common discrete approximation of the full model is the master equation for an effective birth-death process for protein where the propensity of the production reaction is a Hill function of the instantaneous number of proteins whereas the propensity for protein decay is the same as for the usual first-order decay process. Specifically the propensity for the production reaction reads $(L \rho_u + \rho_b n)/(L + n)$ (the symbols as defined in Fig. 1 and above). Hill type propensities of this type or similar are in common use in the literature \cite{hermsen2011speed,aquino2012stochastic,assaf2011determining}. The reduced master equation for this effective birth-death process can be solved exactly in steady-state and it is often thought to be a valid approximation of the full model in the limit of fast promoter switching. It is worth noting that Hill type propensities for protein production are not rigorously derived but rather written by analogy to the effective rate of protein production obtained from the deterministic rate equations under fast equilibrium conditions. Hence the master equation's validity under the same conditions is doubtful \cite{kim2015relationship}. Indeed, it has recently been shown that in the fast switching limit, the steady-state solution of this master equation is precisely the same as that of the Hornos model and hence is not an accurate approximation of the full model if $L$ is small \cite{holehouse2019revisiting} (the solution of this model and that of Hornos are indistinguishable for the parameters in Fig. 2). 

Lastly we consider a recent novel discrete approximation of a class of gene regulatory networks called the Linear Mapping Approximation (LMA \cite{cao2018linear}). In the LMA, in the limit of fast mRNA equilibration, the CME of the full model is approximated by the CME describing bursty protein production and effective promoter switching with no feedback which has an exact steady-state solution. The LMA provides a computational recipe by which the effective rates of promoter switching can be obtained as functions of the parameters in the full model. The LMA together turns out to be the best discrete analytical approximation of the full model, being accurate in all 8 regions of parameters space in Fig. 2. This is followed by Kumar and Grima (both accurate in 4 out of 8 regions) and Hornos (accurate in 2 out of 8 regions). The modified Kumar model does as well as the LMA but it has no known analytical solution. It is to be emphasized that the LMA gives only accurate results when the mean number of proteins conditional on the gene being in state $G$ is much greater than 1, a condition met in all cases considered in Fig. 2.  

In summary, what appear to be minor and subtle differences in the construction of discrete models of auto-regulation, actually lead to considerable differences in the prediction of the steady-state protein distribution. For both positive and negative feedback, the models all agree in only one region of parameter space where the mean burst size and feedback strength are both small (small $B$ and large $L$). The differences between the Grima, Hornos and Kumar models and the full model originate from the fact that these models where not derived rigorously from the full model but rather they were written down intuitively. On the other hand, the LMA does so well because it is derived from the full model. In this section we have considered models of the simplest type of auto-regulatory loop. Discrete models of a loop with more complex mechanisms such as cooperative protein binding to the gene and oscillatory transcription rates, e.g. due to circadian rhythms, have also been solved recently \cite{cao2018linear}. 

\section{Continuous and Hybrid Models of Auto-regulation} 

Besides discrete models there are also continuous models of auto-regulatory loops where it is assumed that molecule number fluctuations correspond to hops on the real axis rather than on an integer axis. The simplicity of the distributions provided by the continuous models often make the results easier to interpret than those obtained from exactly solvable discrete models which are in terms of hypergeometric functions. 

These models are typically described by either the WKB approximation \cite{assaf2011determining,ge2015stochastic,newby2015bistable}, the Linear-Noise Approximation (LNA) (a type of a Fokker-Planck equation) \cite{van1992stochastic,thomas2012slow} or Partial-Integro Differential Equations (PIDEs) \cite{friedman2006linking,pajaro2015shaping,bokes2015protein,jkedrak2016influence}. These are several variations of these three approaches in the literature, too numerous to here enumerate. For the purpose of the present discussion we will compare and discuss two LNA variants and a PIDE approach: (i) by LNA we specifically mean the Fokker-Planck equation obtained by applying the LNA to the CME of the modified Kumar model in Fig 1. (ii) by cLNA we mean the conditional LNA derived in \cite{thomas2014phenotypic} applied to the CME of the modified Kumar model. (iii) by PIDE we specifically mean the model presented by Bokes and Singh for the case that there are no decoy binding sites \cite{bokes2015protein}; its solution is given by Eq. 26 in the SI of the aforementioned paper (this result has been previously reported \cite{friedman2006linking,mackey2013dynamic}). These three have the following different properties: (i) LNA gives a Gaussian distribution, the cLNA gives a sum of two Gaussians, each associated with one of the promoter states while the PIDE gives a unimodal non-Gaussian distribution for the protein numbers (note that generally PIDE can give rise to bimodal distributions but not in this case because there is no cooperativity). (ii) All implicitly take into account mRNA through the protein burst size distribution. This is a geometric distribution for the LNA and cLNA models whereas it is an exponential distribution for the PIDE model. (iii) The LNA and PIDE are continuum models but the cLNA is a hybrid model because protein fluctuations are assumed continuous but gene fluctuations are assumed discrete.   
\begin{figure}[h!]
\centering
\includegraphics[width=1\textwidth]{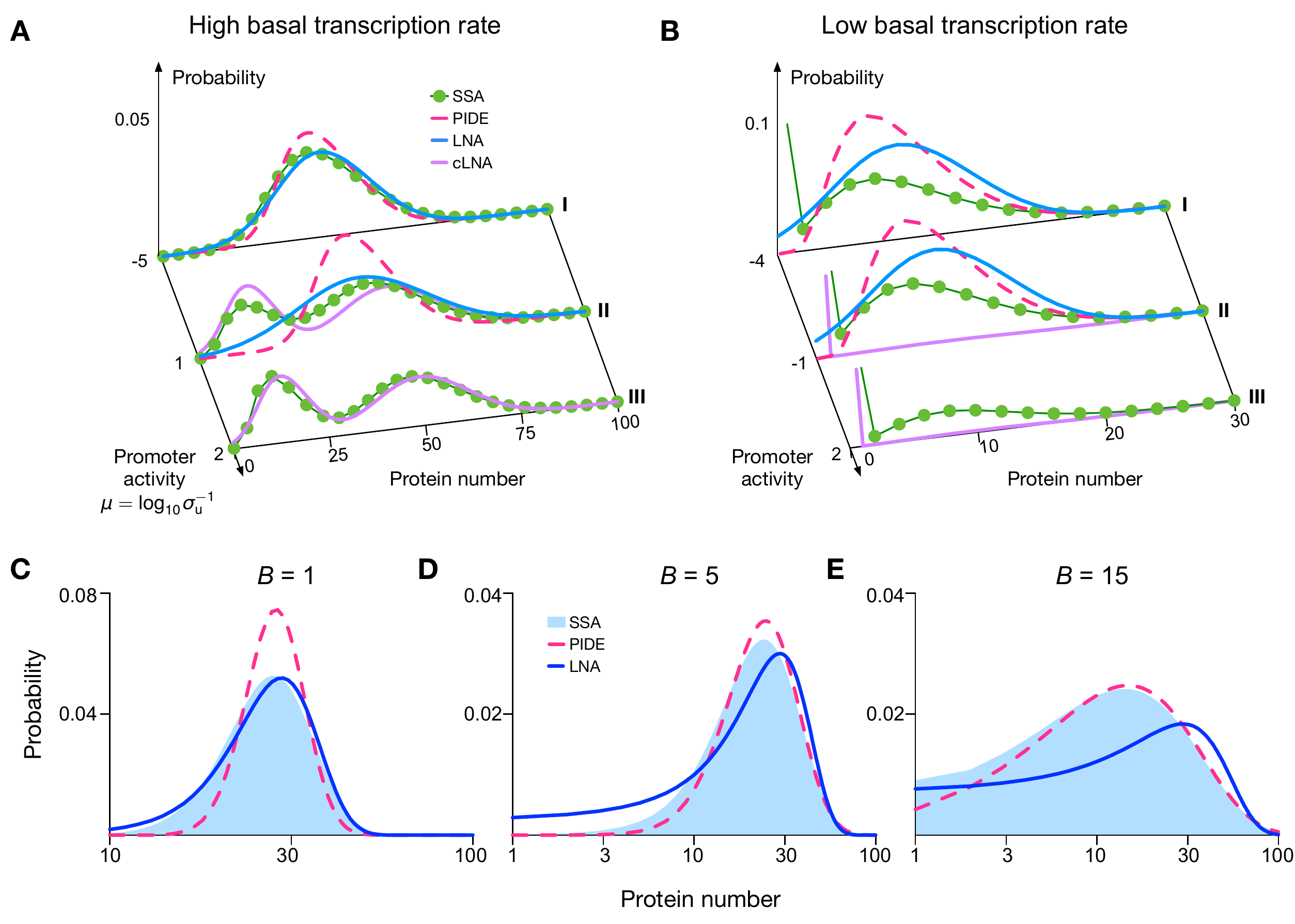}
\caption{(A,B) Plots comparing the accuracy of continuous approximations (LNA, cLNA, PIDE) versus the SSA of the full model under fast, intermediate and slow rates of promoter switching for high and low basal transcription rates. The parameter $\mu$ is large for slow switching and small for fast switching. LNA and PIDE approximations are accurate for fast switching in high basal transcription rate conditions while the cLNA is more accurate for slow switching conditions. All approximations break for low basal transcription. (C,D,E) show plots of LNA vs PIDE vs SSA of full model for fast switching as the mean burst size $B$ increases at constant transcription rates $\rho_u B$ and $\rho_b B$. The LNA performs best at low $B$ while the PIDE performs best at large $B$. Note that the x-axis in (C,D,E) is logarithmic. The parameters are as follows. (A) (I) $\mu = -5$, $\rho_u = 7$, $\rho_b = 25$, $B = 2$, $\sigma_u = 10^5$ and $\sigma_b = 10^4$. (II) $\mu = 1$, $\rho_u = 7$, $\rho_b = 25$, $B = 2$, $\sigma_u = 10^{-1}$ and $\sigma_b = 10^{-2}$. (III) $\mu = 2$, $\rho_u = 7$, $\rho_b = 25$, $B = 2$, $\sigma_u = 10^{-2}$, $\sigma_b = 10^{-3}$. (B) (I) $\mu = -4$, $\rho_u = 10^{-4}$, $\rho_b = 5$, $B = 2$, $\sigma_u = 10^4$ and $\sigma_b = 10^6$. (II) $\mu = -1$, $\rho_u = 10^{-4}$, $\rho_b = 5$, $B = 2$, $\sigma_u = 10$ and $\sigma_b = 10^{3}$ (III) $\mu = 2$, $\rho_u = 10^{-4}$, $\rho_b = 5$, $B = 2$, $\sigma_u = 10^{-2}$ and $\sigma_b = 1$. (C,D,E) $\rho_u B = 30$, $\rho_b B = 20$, $\sigma_u = 10^6$ and $\sigma_b = 10^3$.}
\label{fig3}
\end{figure} 

To summarise our current understanding of the regions of their validity, in Fig. 3A,B we compare the three approximations under fast, intermediate and slow rates of promoter switching for both high and low basal transcription rates ($\rho_u$) versus SSA simulations of the full model. For high basal transcription rates, the LNA and PIDE provide accurate approximations of the full model for fast promoter switching (Fig. 3A, I) and breakdown for intermediate (Fig. 3A, II) and slow promoter switching (Fig. 3A, III) since they cannot capture the bimodality of the full model distribution. In the latter regime, the cLNA performs well instead. These results make sense in the light that the PIDE solution integrates out promoter switching by using a Hill-type function and hence presumes fast promoter switching while the cLNA derivation assumes that promoter switching is much slower than transcription, translation and decay. For intermediate switching, the cLNA misses the precise location of the modes; this is a limitation imposed by trying to fit a Gaussian to each mode which can be corrected for by using the conditional system-size expansion \cite{andreychenko2017distribution}. In contrast, for low basal transcription rate, all three approximations fail independent of the switching rate (Fig. 3B, I-III). This is because continuous approximations are only valid for large enough protein numbers and the low basal transcription rate induces a mode of the protein distribution at zero. In Fig. 3C, D, E we compare the LNA and PIDE approximations as the mean burst size $B$ is increased at constant transcription rate for fast switching conditions. The LNA does best for low burst sizes because the full model distribution is almost Gaussian; the PIDE approximation is inaccurate here because for small mean burst sizes, the exponential distribution of burst sizes is not a good approximation of the geometric distribution. The reverse is true when the mean burst size becomes large: the full model distribution is highly non-Gaussian and cannot be captured by the LNA but is well captured by the PIDE, also because the exponential is a good approximation of the geometric distribution in this case. 

In summary, various continuum models provide reasonably accurate analytical steady-state distributions for protein numbers in terms of simple functions, provided there is not a mode of the distribution at zero and provided one is in the limit of either fast or slow switching. In the literature, there are also analytically solvable continuum models with multiple gene copies \cite{jkedrak2016influence} and a few results are also known for promoter switching that is neither extremely slow nor exceedingly rapid \cite{ge2015stochastic}. 

\section{Insights from models}

There are at least two main insights obtained from the analysis of stochastic models of auto-regulation: (i) Cooperativity is not necessary for protein distributions to be bimodal. Slow or intermediate promoter switching can in some cases lead to bimodality, in the absence of cooperativity. (ii) There is not a simple relationship between noise reduction or amplification and the type of feedback loop (negative or positive). We next discuss each of these in detail. 

\textit{Insight}(i). An important property of auto-regulatory feedback loops is their ability to generate protein distributions with more than one mode. Each of these modes can be associated with a sub population of cells of a particular phenotype and hence their quantification is important for understanding cellular decision-making. These modes can arise in at least two ways: (a) If the deterministic rate equations are bistable (typically arising from cooperativity) and provided leakage remains below a critical threshold \cite{pajaro2015shaping}; (b) If the deterministic rate equations are not bistable and there is substantial noise due to the switching of a molecule between a  discrete number  of conformational  states. This last case is also called noise-induced bistability \cite{qian2009stochastic}. Hence bimodality in protein distributions does not require cooperativity \cite{ochab2010bimodal}. If promoter switching is slow enough then the system will alternate between two steady-state protein distributions, each associated with a promoter state; hence if these distributions are well separated then the full distribution will appear to be bimodal and we can then say that ``noise in the gene states created the bistability''. This has also been shown to lead to birhythmical expression in genetic oscillators and to hysteresis in phenotypic induction  \cite{thomas2014phenotypic}; furthermore it leads to an enhancement of the sensitivity of the circuit's response to input signals \cite{hermsen2011speed}. The LNA cannot capture either type of bimodality because it is valid for those systems whose deterministic rate equations are monostable \cite{schnoerr2017approximation}, provided the average number of molecule numbers is large enough. Hybrid methods such as the cLNA and others \cite{kurasov2018stochastic,karmakar2007positive,lin2016gene} can capture bimodality of type (b) because they model gene states discretely. Methods based on PIDE can capture bimodality of type (a) but not (b) since they do not model discrete gene states explicitly \cite{pajaro2015shaping}. The same is true for protein distributions obtained using the Fokker-Planck equation stemming from the Kramers-Moyal expansion \cite{duncan2015noise}. Discrete models can capture both types of bimodality \cite{cao2018linear}.

\parindent 5mm \textit{Insight} (ii). Early work reported that negative feedback reduces protein fluctuations \cite{simpson2003frequency} whereas positive feedback has the opposite effect \cite{hasty2000noise}. For negative feedback there is an optimal feedback strength at which the protein fluctuations are minimal \cite{singh2009optimal}. However later work showed that models with different assumptions can yield contradictory conclusions about how feedback affects noise \cite{marquez2010counter}. More recently it has been claimed that the effect of feedback on noise can be more easily understood via a noise decomposition \cite{liu2016decomposition}. Noise can be decomposed as the sum of three types: promoter noise, birth-death noise, and correlation noise induced by feedback. In the case of slow switching, where the promoter noise is dominant, positive feedback reduces the total noise, whereas negative feedback amplifies it. In the case of fast switching, where the correlation noise is dominant, positive feedback amplifies the total noise, whereas negative feedback reduces it. Further work in this direction can be found here  \cite{jia2017stochastic,jia2019macroscopic}. Hence it appears that the general intuition that negative feedback reduces noise while positive feedback increases it, is not correct. It follows that the ubiquity of negative self-regulating transcription factors in prokaryotic cells cannot be explained by an evolutionary pressure to select for mechanisms that enable control of protein noise \cite{stekel2008strong}. 

\section{Open problems}

In summary, our literature review and comparative analysis shows that subtle differences between models of auto-regulation can have a significant impact on the predicted distribution of protein numbers, e.g. different number of modes and different predictions for how positive and negative feedback influence the size of protein number fluctuations. The current generation of reduced models have been constructed intuitively (not derived rigorously from an underlying fine-scale model) and hence the existing differences between them.

We conclude by briefly pointing out a few of the open problems in the field: (i) The exact solution of the discrete modified Kumar model. The motivation for its study stems from the fact that stochastic simulations show that (unlike the Grima, Hornos and Kumar models in the literature) it is in excellent agreement with the full model for all feedback strengths and protein burst sizes, provided mRNA decays much quicker than protein. (ii) The derivation of exact time-dependent solutions of the CME for the Grima, Kumar and modified Kumar models (it is presently known for the Hornos model \cite{ramos2011exact}). The derivation of approximate time-dependent propagators has received recent attention \cite{veerman2018time}. Explicit time-dependent solutions would enable a detailed study of how perturbations to an auto-regulatory circuit, e.g. inhibition of certain reactions, affects the dynamics -- this would aid the interpretation of experiments of this type. (iii) The development of new continuous approximation methods that can accurately predict the steady-state distribution of protein number of the full model without the assumption of fast or slow promoter switching, i.e. valid for any intermediate switching. (iv) The derivation of reduced models of feedback loops starting from fine-grained stochastic models incorporating biological processes which are known to affect gene expression such as partitioning of proteins due to cell division, gene replication, gene dosage compensation and nascent mRNA maturation (such models have been studied using simulations for systems with no feedback \cite{skinner2016single}). Likely such reduction is possible by the careful application of timescale separation methods or possibly by mapping to an effective CME with stochastic rates of transcription, translation and feedback \cite{park2018chemical}. 

In our opinion, the last of these open problems is probably the hardest and the most pressing since it is imperative that the reduced models studied are biologically and physically realistic before further mathematical analysis is undertaken. 

\section*{Acknowledgments}{This work was supported by a BBSRC EASTBIO PhD studentship, BBSRC grant BB/M025551/1 and the UK Research Councils' Synthetic Biology for Growth programme of the BBSRC, EPSRC and MRC (BB/M018040/1).}

\bibliographystyle{unsrt}
\bibliography{biblio}
\end{document}